\documentclass[11pt]{article}
\usepackage{amsmath,amssymb,amsthm}

\usepackage{graphicx}
\def \ds {\displaystyle}

\def \ns {\normalsize}

\def \es {\enspace}
\def \ts {\thinspace}
\def \nts {\negthinspace}

\title{
  \vspace*{-1.5cm}
  \hfill{\ns KEK-TH-1346} \\ 
  \vspace*{-0.3cm}
  \hfill{\ns January 2010}  \\
  \vspace*{3.0cm}
  {\Large {\bf Matrix Model and Elliptic Curve}} \\
  \vspace*{0.8cm}
}
\author{ Hirotaka Sugawara\thanks{e-mail: hirotaka.sugawara@kek.jp} }
\date{}
\begin{document}
\maketitle
\begin{center}
{\it JSPS Washington Office, 1800 K Street NW \#920, Washington, DC 20006,} \\
{\it Department of Physics and Astronomy, Johns Hopkins University,} \\ 
{\it and} \\
{\it High Energy Accelerator Research Organization (KEK),
Tsukuba, Ibaraki, Japan} \\
\end{center}

\vspace{6cm}
%
%
\begin{abstract}
Solution to the reduced matrix model of IKKT type is studied with non-zero fermion
fields. A suggestion is made that our universe is made of rational numbers rather
than being a continuum. To substantiate this proposal, the reduced Yang-Mills
equation is written in the form of an elliptic curve. The normalization of the 
solution can be expressed in terms of the Weierstrass function generically or
in terms of the Dedekind function in the case of 3-brane. A way to define the 
gravitational field in the matrix model is proposed with some new interpretation
of the cosmological constant. The (first) quantization of the system is done
within the framework of non-commutative geometry.
\end{abstract}
\clearpage
%
\section{Introduction}
Some time ago, two groups, independently of each other, 
(ref.\,\cite{BFFS,IKKT}) made a proposal to formulate the string theory in
terms of matrix models. One\,\cite{BFFS} deals with the M-theory in the light
cone gauge and the other\,\cite{IKKT}) defines the IIB string theory in terms
of the reduced matrix model\,\cite{EK}. Here, I follow the latter formulation
but without being restricted to the IIB model.

  The purpose of this paper is to point out a simple fact that the matrix model
can be considered within the framework of number theory in a rather elementary
way. The motivation of the study is the following.

  First, the number of the matrix elements of the matrix models will be
infinite but obviously it is not continuous. A conventional wisdom is to take
some continuous limit of the matrix model such as to take $N \to \infty$
and $R \to \infty$ with $N/R$ fixed to obtain the desired space time continuum.
If we take the matrix model as it is as the fundamental theory of space-time
without taking the continuum limit, a new insight into the problem emerges.
Space-time is composed of integers or at most rational numbers if it is a
dense set, as is usually conceived. To give a substance to this view, we give
a new meaning to the reduced Yang-Mills equation. Crudely speaking,
the equation reads,
\begin{eqnarray}
A \otimes A \otimes A = \psi \times \psi \ts.
\label{eq:triProduct}
\end{eqnarray}
This is reminiscent of the so-called elliptic curve\,\cite{MP} in the number
theory which can be written in general as
\begin{eqnarray}
a x^3 + b x^2 + c x + d = y^2 \ts.
\label{eq:ellipticCurve}
\end{eqnarray}
  It is important that we retain the fermion term (classical De Broglie field)
in (\ref{eq:triProduct}) to make a comparison. Since $A$ and $\psi$ in
(\ref{eq:triProduct}) are
infinite matrices with the product symbols having a specific meaning, we are
treating here an infinite set of coupled elliptic curves as  will be shown later
in this paper.  We assert that the coefficients of these equations are integers
or rational numbers and the solutions are also rational numbers. This view
enables us to utilize the whole machineries of number theory developed over
the last centuries.

  For example, we are interested in the number of solutions to equation
(\ref{eq:ellipticCurve}). A zeta function corresponding to equation
(\ref{eq:ellipticCurve}) can be defined to study this problem. In the
simple case of integer coefficients, this can be defined as;
\begin{eqnarray}
\zeta(z) = \prod_p \left( 1 - a_p p^{-z} + p^{1-2z} \right)^{-1} \ts,
\label{eq:zetaFunc}
\end{eqnarray}
where $a_p = p - N_p$ with $N_p$ the number of solutions of
(\ref{eq:ellipticCurve}) between $0$ and $p$. The product is over all the
prime numbers $p$ except for those $p$ which give rise to a multiple root
when (\ref{eq:ellipticCurve}) is considered modulo $p$.

In the case of equation (\ref{eq:triProduct}), we assume that the infinite
matrix A is composed of clusters of finite matrices and we define zeta
function for each of them. If the zeta function for a certain cluster
satisfies $\zeta(1) = 0$, we see that we have infinite number of rational
number solutions to this cluster due to Birch and Swinnerton-Dyer
conjecture\,\cite{MP}. This is almost needed since we have to solve equation
(\ref{eq:triProduct}) together with the other equation of motion which
looks symbolically as $A \otimes \psi = 0$. As is well known the elliptic
curve (\ref{eq:ellipticCurve}) is nothing but a two torus and the relation of
$(x,\ts y)$ to the complex coordinate on the upper half plane is given by;
\begin{eqnarray}
x = \wp(z) \ts,\es y = \frac{\ds 1}{\ds 2} \wp^{\prime}(z) \ts,
\quad \mbox{with} \es \wp^{\prime}(z)^2 
= 4\wp(z)^3 - g_2 \wp(z) - g_3 \ts,
\label{eq:WeierstrassFunc}
\end{eqnarray}
where $\wp(z)$ stands for the Weierstrass function. The elliptic curve
appeared in physics in different context. Notably, in the analysis of
modular space of $N=2$ supersymmetric gauge theory (Seiberg and
Witten\,\cite{SeiWitten}), the vacuum values of the fields are defined on
the elliptic curve with coordinate of the moduli space u being the zero of
$\wp^{\prime}(z)$. There exist many papers which deal with elliptic curves
in various contexts of physics but these are not referred to here since
they are not relevant to the present work\,\cite{MORE3DOZ}. In obtaining
the zeta function for the equation (\ref{eq:triProduct}) the Shimura
conjecture (or Taniyama-Shimura conjecture)\,\cite{SHTA} might be helpful.
This famous conjecture was proven by A.~Wiles\,\cite{Wiles} in a special
case which is needed for the proof of Fermat's theorem and in its full
generality by Ch.~Breuil, B.~Conrad, F.~Diamond and R.~taylor in
1999\,\cite{FCDRT}. The conjecture says that the zeta function
(\ref{eq:zetaFunc}) is modular.

  In this context, what I am trying to propose in this paper is to assert
that equation (\ref{eq:triProduct}) should be considered as modular curve.
This naturally leads to the space-time which is composed of field of
rational numbers or its algebraic extension.

  One of the main research goals of number theory is to generalize the
Shimura conjecture to the zeta functions of higher degrees and with
an extended automorphism. This so-called Langlands
program\,\cite{FCDRT} has a geometric counter part and is being
extensively studied in connection with string theory or chiral
theory\,\cite{KapWitten,Frenkel}.

  The organization of the paper is the following:
Section\,\ref{sec:Preparations} prepares some necessary machinery
such as the description of the fermion de Broglie fields. The cluster of
the matrix will also be defined. Section\,\ref{sec:EOM} is devoted to
the analysis of equation (\ref{eq:triProduct}) with the proof that
the scale of the classical fields is given by the Weierstrass function or
by the Dedekind function. In section\,\ref{sec:Gravity}, a method is
proposed to introduce the gravitation into the matrix models.
Section\,\ref{sec:1stQiztnNCG} is devoted to the problem of (first)
quantization. The conjugate variables are introduced as the Dirac
operators of the non-commutative geometry.
Section\,\ref{sec:Conclusions} is for the brief concluding remarks.
\vspace*{5mm}
%
\section{Some Preparations}\label{sec:Preparations}
The action of the matrix model is written in such a way to accommodate
the non-associative case:
\begin{eqnarray}
L = - \frac{\ds 1}{\ds 4} \mbox{Tr}
\left\{ [A_{\mu},A^{\nu}] [A^{\mu},A_{\nu}] \right\} 
- \frac{\ds g}{\ds 2} \mbox{Tr} \left\{ \overline{\Psi}
(A_{\mu}\Gamma^{\mu}\Psi) - (A_{\mu}\overline{\Psi}\Gamma^{\mu})
\Psi \right\} \ts,
\label{eq:actionIKKT}
\end{eqnarray}
where $A_{\mu}$ is a 10-dimensional Minkowski vector and $\Psi$ is
a 10-dimensional 32-component spinor. Depending on whether we take the
IIB case or the IIA case, the expression for the $\Psi$ is different.

For the IIB case, we have,
\begin{eqnarray}
\Psi = \left\{ (x + x_{a,b}\Gamma^{a+}\Gamma^{b+} + yB_1)
+ \Gamma^{0+} (x_a\Gamma^{a+} + y_a\Gamma^{a+}B_1) \right\} \zeta \ts.
\label{eq:Psi_IIB}
\end{eqnarray}
Here the notation is that of Polchinski\,\cite{Polchinski}:
\begin{eqnarray*}
\Gamma^{0\pm}
\nts&=&\nts \frac{\ds 1}{\ds 2} (\pm\Gamma^{0} + \Gamma^{1}) \ts,\es
\Gamma^{a\pm} = \frac{\ds 1}{\ds 2} (\Gamma^{2a} \pm \Gamma^{2a+1}) \ts,
\quad (a =1,2,3,4) \\
B_1
\nts&=&\nts \Gamma^{3}\Gamma^{5}\Gamma^{7}\Gamma^{9} \ts, \\
\Gamma^{a+}
\nts& &\hspace*{-7mm} \mbox{are creation operators and } \ts
\Gamma^{a-} \es\mbox{are annihilation operators including } a=0 \ts, \\
\zeta 
\nts& &\hspace*{-7mm} \mbox{is a ground state with } 
\Gamma^{a+} \zeta = 0 \es\mbox{for} \es a=0,1,2,3,4 \ts, \\
\nts& &\hspace*{-17mm} 
x,\ts x_{a,b},\ts y,\ts x_a \es\mbox{and}\es y_a 
\es\mbox{are all complex numbers.}
\label{def:GammaB_IIB}
\end{eqnarray*}
This case corresponds to two 16-component Majorana-Weyl spinors of
same chirality.

For the IIA case we have similarity:
\begin{eqnarray}
\Psi 
\nts\nts&=&\nts\nts \bigl\{ (x + x_{a,b}\Gamma^{a+}\Gamma^{b+} + yB_1)
+ \Gamma^{0+} (x_a\Gamma^{a+} + y_a\Gamma^{a+}B_1) \nonumber \\
\nts\nts&+&\nts\nts \Gamma^{0+}
(x^{\prime} + x_{a,b}^{\prime}\Gamma^{a+}\Gamma^{b+} + y^{\prime}B_1)
+ (x_{a}^{\prime}\Gamma^{a+} + y_{a}^{\prime}\Gamma^{a+}B)
\bigr\} \ts \zeta \es.
\label{eq:Psi_IIA}
\end{eqnarray}
The Majorana condition reads,
\begin{eqnarray*}
& &\hspace*{-10mm}
x^{\ast} = y,\ts
x_{a,b}^{\ast} = - \varepsilon_{abcd} y_{c,d},\ts
x_{a}^{\ast} = y_{a}, \\
& &\hspace*{-11mm}
x^{\prime\ast} = y^{\prime},\ts
x_{a,b}^{\prime\ast} = - \varepsilon_{abcd} y_{c,d}^{\prime},\ts
x_{a}^{\prime\ast} = y_{a}^{\prime} \es\mbox{and}\es
\zeta^{\ast} = \zeta \ts.
\label{def:Majorana_IIA}
\end{eqnarray*}
We have two Majorana-Weyl spinors of opposite chirality in this case.
The variables $x's$ and $y's$ are all infinite compoment matrices as 
is $A_{\mu}$.

The next task is to define the cluster and to express the infinite
matrix as:
\begin{equation}
A_{\mu} = 
\left(
 \begin{array}{ccccc}
 A_{\mu}^{(1)} &  &  &  &  \\
   & A_{\mu}^{(2)} &  &  &  \\
   &  & A_{\mu}^{(3)} &  &  \\
   &  &  & \cdots & \\
   &  &  &  & \cdots
 \end{array}
\right)
\label{eq:SL2Z}
\end{equation}
We assume that each finite dimensional cluster matrix
$(A_{\mu},\ts x,\ts y,\ldots)$ to be expanded as (suppressing the cluster
index):
\begin{equation*}
A_{\mu} = A_{i,\mu} \, \lambda_{i} \ts,
\label{eq:matrix_expnd}
\end{equation*}
where $\lambda_{i}$ satisfies the following multiplication rules:
\begin{equation}
\lambda_{i} \lambda_{j} = \, (i f_{ijk} + d_{ijk}) \lambda_{k}
+ (g_{ij} + i e_{ij}) I \es.
\label{eq:lambda_multiplication}
\end{equation}
The only conditions for the real parameters $f, d, g$ and $e$ are that
$f$ and $e$ are antisymmetric with respect to the interchange of the first
two suffices and $d$ and $g$ are symmetric. In fact, the reality of 
$\overline{\Psi} \Gamma^{\mu} \Psi$ imposes further condition that either
of the following two cases be satisfied,
\begin{itemize}
\setlength{\itemsep}{-2pt}
\item[(1)] $e_{ij} = 0$, $d_{ijk}$ is cyclic symmetric in $i,j,k\ts $,
\item[(2)] $e \ne 0,\ts d = 0\ts $.
\end{itemize}
Coexistence of $f$ and $e$ (case (2)) corresponds to the non-associative
case. In this case, the matrix representation exists using the non-associative algebra itself as the representation space but the multiplication rule
(\ref{eq:lambda_multiplication}) should be applied rather than the regular
matrix rule. We also assume, $T_{r}(\lambda_{i}) = 0,\ts
T_{\tau}(I) = a$ (some finite number).

The motivation to generalize the matrix model to the non-associative case
stems from the recent interest in the non-associative algebra which arose
in connection with the M2-brane analysis initiated by J.~Bagger and
N.~Lambert\,\cite{JBNL}.

The simplest non-associative example of equation
(\ref{eq:lambda_multiplication}) is given by the following
``generalized Pauli spin":
\begin{eqnarray}
\sigma_{i} \sigma_{j} = i \varepsilon_{ijk} \sigma_{k} +
(\delta_{ij} + i \kappa \varepsilon_{ij}) I \ts.
\label{def:gnlzdPauliSpin}
\end{eqnarray}
For $x = a_{i}\sigma_{i} + a_{0} I,\ts y = b_{i}\sigma_{i}
 + b_{0} I,\ts z = c_{i}\sigma_{i} + c_{0} I$, we have,
\begin{itemize}
\setlength{\itemsep}{-2pt}
\item[(1)] Jordan condition: $(xy)z - x(yz) = 0$,
\item[(2)] symmetrized associator: $[ x,\ts y,\ts z ] 
= 4i\kappa \left\{ (\vec{a}\times\vec{b}\cdot\vec{I})\vec{c} 
+ (\vec{b}\times\vec{c}\cdot\vec{I})\vec{a}
+ (\vec{c}\times\vec{a}\cdot\vec{I})\vec{b}
\right\} \,\vec{\sigma} \ts,$
\end{itemize}
where $\vec{I}$ is a vector with a unit length of all three components
and $\vec{a} = (a_1,\ts a_2,\ts a_3)$ etc.

Since we regard the each cluster to correspond to a small portion of
space-time coordinate, the order of assigning each solution to a cluster
should not matter. This will be regarded as the transformation which
replaces the general coordinate transformation as will be discussed
in section\,\ref{sec:Gravity}.
\vspace*{5mm}
%
\section{Equations of Motion}\label{sec:EOM}
  From the action (\ref{eq:actionIKKT}) we get the following
equations of motion for each cluster:
\begin{eqnarray}
& &\hspace*{-10mm}
4G_{ij;i^{\prime}j^{\prime}} A^{\nu}_{j} A^{\mu}_{i^{\prime}}
A_{j^{\prime}\nu} - \frac{g}{2} F_{jik} 
\overline{\Psi}_{j} \Gamma^{\mu} \Psi_{k} = 0 \ts, 
\label{eq:eom_clusters_1} \\
& &\hspace*{-10mm}
F_{ijk} A_{j\mu} \Gamma^{\mu} \Psi_{k}  = 0 \ts,
\label{eq:eom_clusters_2}
\end{eqnarray}
where
\begin{eqnarray}
& &\hspace*{-10mm}
G_{ij;i^{\prime}j^{\prime}} = f_{ijk}f_{i^{\prime}j^{\prime}k^{\prime}}
g_{kk^{\prime}} + \kappa^2 e_{ij}e_{i^{\prime}j^{\prime}} \ts, \\
& &\hspace*{-10mm}
F_{jik} = (if_{jkl} + d_{jkl})(g_{il} + i \kappa e_{il}) 
- (if_{jil} + d_{jil})(g_{lk} + i \kappa e_{lk}) \ts,
\label{def:eom_clusters}
\end{eqnarray}
%
We can write
\begin{eqnarray}
F_{ijk} = i \tilde{f}_{ijk} + \tilde{d}_{ijk} \ts,
\label{def:F_ijk}
\end{eqnarray}
Then the condition that the fermion term in the equation
(\ref{eq:eom_clusters_1}) be real is satisfied by $d$ or $f$ being symmetric
or anti-symmetric respectively when we exchange the first and the last
suffices and this leads to the condition stated under the equation
(\ref{eq:lambda_multiplication}).
We define following variables:
\begin{eqnarray}
A_{j,2c} + i A_{j,2c+1} \equiv \alpha_{j,c} 
= |\alpha_{c}| \beta_{j,c} \ts,
\label{def:variables}
\end{eqnarray}
with
\begin{eqnarray}
\sum |\beta_{j,c}|^2 = 1 \ts,
\label{def:beta_jc}
\end{eqnarray}
and
\begin{eqnarray}
A_{j,1} \pm A_{j,0} = \alpha_{j,0}^{\pm}
= |\alpha_{0}| \beta_{j,0}^{\pm} \ts,
\label{def:A_alpha_j0}
\end{eqnarray}
with
\begin{eqnarray}
\beta_{j,0}^{+} \beta_{j,0}^{-} = \pm 1 \es (+ \ts\mbox{for space-like and}
\ts -\ts\mbox{for time-like vectot} \es A_{\mu}) \ts.
\label{def:beta_j0}
\end{eqnarray}

Then the equation (\ref{eq:eom_clusters_2}) can be reduced to the
following five equations (we write down the IIB case):
\begin{eqnarray}
& &\hspace*{-10mm}
F_{ijk} (\alpha_{j,a}x_{k,a} - \alpha_{j,0}^{+}x_{k}) = 0 \ts,
\label{eq:F_ijk-alpha_1} \\
& &\hspace*{-10mm}
F_{ijk} (\alpha_{j,a}^{\ast}y_{k,a} - \alpha_{j,0}^{+}y_{k}) = 0 \ts,
\label{eq:F_ijk-alpha_2} \\
& &\hspace*{-10mm}
F_{ijk} \bigl\{ \frac{1}{2}
(\alpha_{j,a}^{\ast}y_{k,b} - \alpha_{j,b}^{\ast}x_{k,a}) -
\alpha_{j,0}^{+}x_{k;a,b} - \varepsilon_{abcd} \alpha_{j,c}y_{k,d}
\bigr\} = 0 \ts,
\label{eq:F_ijk-alpha_3} \\
& &\hspace*{-10mm}
F_{ijk} (\alpha_{j,a}y_{k} + \alpha_{j,0}^{-}y_{k,a}
+ \varepsilon_{abcd} \alpha_{j,b}^{\ast}x_{k;c,d}) = 0 \ts,
\label{eq:F_ijk-alpha_4} \\
& &\hspace*{-10mm}
F_{ijk} (\alpha_{j,a}^{\ast}x_{k} - 2\alpha_{j,b}x_{k;a,b}
+ \alpha_{j,0}^{-}x_{k,a}) = 0 \ts,
\label{eq:F_ijk-alpha_5} 
\end{eqnarray}
Here the repeated suffices must be summed over.
(\ref{eq:eom_clusters_1}) can be written as (again IIB case)
\begin{eqnarray}
4G_{ij;i^{\prime}j^{\prime}} A^{\nu}_{j} A^{\mu}_{i^{\prime}}
A_{j^{\prime}\nu} - \frac{g}{2} F_{jik}
\langle \Gamma^{\mu} \rangle_{j,k} = 0 \ts,
\label{eq:VEV_eom_clusters_1}
\end{eqnarray}
with
\begin{eqnarray}
\langle \Gamma^{0} \rangle_{j,k}
\nts\nts&=&\nts\nts x^{\ast}_{j}x_{k} 
+ 2 x^{\ast}_{j;a,b}x_{k;a,b} + y^{\ast}_{j}y_{k} 
+ x^{\ast}_{j,a}x_{k,a} + y^{\ast}_{j,a}y_{k,a} \ts,
\label{eq:VEV_Gamma_0} \\
\langle \Gamma^{1} \rangle_{j,k}
\nts\nts&=&\nts\nts x^{\ast}_{j}x_{k} 
+ 2 x^{\ast}_{j;a,b}x_{k;a,b} - \left( y^{\ast}_{j}y_{k} 
+ x^{\ast}_{j,a}x_{k,a} + y^{\ast}_{j,a}y_{k,a} \right) \ts,
\label{eq:VEV_Gamma_1} \\
\langle \Gamma^{2a} \rangle_{j,k}
\nts\nts&=&\nts\nts - x^{\ast}_{j}x_{k,a} - x^{\ast}_{j,a}x_{k}
- y^{\ast}_{j}y_{k,a} - y^{\ast}_{j,a}y_{k} \nonumber \\
\nts\nts&-&\nts\nts\nts 2 x^{\ast}_{j;a,b}x_{k,b}
- 2 x^{\ast}_{j,b}x_{k;a,b} - \varepsilon_{abcd}
\left( x^{\ast}_{j;c,b}y_{k,d} + y^{\ast}_{j,d}x_{k;c,b} \right) \ts,
\label{eq:VEV_Gamma_2a} \\
\langle \Gamma^{2a+1} \rangle_{j,k}
\nts\nts&=&\nts\nts -i\bigl\{ x^{\ast}_{j}x_{k,a} - x^{\ast}_{j,a}x_{k}
- y^{\ast}_{j}y_{k,a} + y^{\ast}_{j,a}y_{k} \nonumber \\
\nts\nts&-&\nts\nts\nts 2 x^{\ast}_{j;a,b}x_{k,b}
+ 2 x^{\ast}_{j,b}x_{k;a,b} - \varepsilon_{abcd}
\left( - x^{\ast}_{j;c,b}y_{k,d} 
+ y^{\ast}_{j,d}x_{k;c,b} \right) \bigr\} \ts,
\label{eq:VEV_Gamma_2a+1}
\end{eqnarray}
In writing down these equations, the Grassmannian factor
$\zeta^{\ast}\zeta$ or $\zeta^{\rm T}\zeta$ (in case of IIA) is absorbed
in the coupling constant $g$. In fact, the easiest way to get rid of
Grassmannian factor is to assume that all $A^{\mu}$
contain a factor $(\zeta^{\ast}\zeta)^{2/3}$ (IIB) or 
$(\zeta^{\rm T}\zeta)^{2/3}$ (IIA). Finally the equation
(\ref{eq:VEV_eom_clusters_1}) will become the following
two sets of equations:
\begin{eqnarray}
& &\hspace*{-10mm} \rho_{i}^{\pm} |\alpha_{0}|^{3} + 
\bigl( \sum_{a}|\alpha_{a}|^{2} \sigma_{i,a}^{\pm} \bigr) |\alpha_{0}|
= \frac{g}{4} \gamma_{i}^{0\pm} |\lambda|^{2} \ts, 
\label{eq:dfm_eom_clusters_1} \\
& &\hspace*{-10mm} \sigma_{i;c,c} |\alpha_{c}|^{3}
+ \bigl( \sum_{a \ne c} \sigma_{i;a,c} |\alpha_{a}|^{2}
+ \rho_{i,c} |\alpha_{0}|^{2} \bigr) |\alpha_{c}|
= \frac{g}{4} \gamma_{i}^{c} |\lambda|^{2} \ts,
\label{eq:dfm_eom_clusters_2}
\end{eqnarray}
where
\begin{eqnarray}
\gamma_{i}^{0\pm}
\nts\nts&=&\nts\nts 
\tilde{\gamma}_{i}^{1} \pm \tilde{\gamma}_{i}^{0} \ts, 
\label{def:gamma_pc_0pm} \\
\gamma_{i}^{c}
\nts\nts&=&\nts\nts 
\tilde{\gamma}_{i}^{2c} + i \tilde{\gamma}_{i}^{2c+1} \ts,
\label{def:gamma_pc_c}
\end{eqnarray}
with
\begin{eqnarray}
F_{jik} \langle \Gamma^{\mu}_{j,k} \rangle 
= \tilde{\gamma}_{i}^{\mu} |\lambda|^{2} \ts,
\label{def:F_ijk_VEV_gamma}
\end{eqnarray}
and
\begin{eqnarray}
\rho_{i}^{\pm}
\nts\nts&=&\nts\nts G_{ij;i^{\prime}j^{\prime}}
\bigl( \beta_{j,0}^{+}\beta_{j^{\prime},0}^{-} 
+ \beta_{j,0}^{-}\beta_{j^{\prime},0}^{+} \bigr)
\beta_{i^{\prime},0}^{\pm} \ts, 
\label{def:rho_i_pm} \\
\sigma_{i,a}^{\pm}
\nts\nts&=&\nts\nts G_{ij;i^{\prime}j^{\prime}}
\bigl( \beta_{j,a}\beta_{j^{\prime},a}^{\ast} 
+ \beta_{j,a}^{\ast}\beta_{j^{\prime},a} \bigr)
\beta_{i^{\prime},0}^{\pm} \ts, 
\label{def:sigma_ia_pm} \\
\rho_{i,c}
\nts\nts&=&\nts\nts G_{ij;i^{\prime}j^{\prime}}
\bigl( \beta_{j,0}^{+}\beta_{j^{\prime},0}^{-} 
+ \beta_{j,0}^{-}\beta_{j^{\prime},0}^{+} \bigr)
\beta_{i^{\prime},c} \ts, 
\label{def:rho_ic} \\
\sigma_{i;a,c}
\nts\nts&=&\nts\nts G_{ij;i^{\prime}j^{\prime}}
\bigl( \beta_{j,a}\beta_{j^{\prime},a}^{\ast} 
+ \beta_{j,a}^{\ast}\beta_{j^{\prime},a} \bigr)
\beta_{i^{\prime},c} \ts, 
\label{def:sigma_iac}
\end{eqnarray}
The equations (\ref{eq:dfm_eom_clusters_1}) and
(\ref{eq:dfm_eom_clusters_2}) have the desired form of elliptic curves.
They must give the same value of $|\alpha_{0}|$ or $|\alpha_{0}|$ for
the $i$-dependent coefficients. The consistency condition can be obtained
in the following way: We first solve the equations for some value of $i$
and  substitute this back to equations (\ref{eq:dfm_eom_clusters_1})
and (\ref{eq:dfm_eom_clusters_2}). Then we have
\begin{eqnarray}
|\alpha_{0}| 
\nts\nts\nts&=&\nts\nts\nts \wp(\omega) \quad \mbox{with}
\es \wp^{\prime}(\omega) = 0 \ts,
\label{eq:wp_zero1} \\
            & &\hspace*{-1.4cm} \mbox{{\it and}}  \nonumber \\ 
|\alpha_{c}| 
\nts\nts\nts&=&\nts\nts\nts \wp_{c}(\omega_{c}) \quad \mbox{with}
\es \wp^{\prime}(\omega_{c}) = 0 \ts,
\label{eq:wp_zero2} 
\end{eqnarray}
where $\wp$ and $\wp_{c}$ are Weierstrass functions. The equations
(\ref{eq:dfm_eom_clusters_1}) and (\ref{eq:dfm_eom_clusters_2})
become respectively,
\begin{eqnarray}
& & 4\wp^{3}(\omega) \rho_{i}^{\pm} + 4\sum_{c} \wp_{c}^{2}(\omega_{c}) 
\sigma_{i,c}^{\pm} \wp(\omega) = g\gamma_{i}^{\pm} \ts,
\label{eq:dfm_eom_clt_1} \\
& & 4\sum_{c} \wp_{c}^{3}(\omega_{c})\sigma_{i;c,a}
+ 4\wp^{2}(\omega) \wp_{a}(\omega_{a}) \rho_{i,a} = g\gamma_{i}^{a} \ts,
\label{eq:dfm_eom_clt_2} 
\end{eqnarray}
These two equations are regarded as the consistency conditions but they
themselves are a kind of coupled elliptic curves. My proposal is to define
all these equations (coefficients and the solutions) on the field of rational 
numbers. The complex numbers must have real and imaginary parts which are
rational respectively.

The 3-brane case can be further reduced to a simpler form as follows:
This case is defined by 
\begin{eqnarray*}
& & A^{\mu} \ne 0 \quad \mbox{only for} \es \mu=0,1,2,3 \ts, \\
& & x_{j;a,b} = 0 \ts, \\
& & x_{j,a} \ne 0 \quad \mbox{only for} \es a=1 \ts, \\
& & y_{j,a} \ne 0 \quad \mbox{only for} \es a=1 \ts, \\
& & x_{j} \ne 0 \ts, \es y_{j} \ne 0
\end{eqnarray*}
The equations (\ref{eq:F_ijk-alpha_1}) to (\ref{eq:F_ijk-alpha_5}) become,
\begin{eqnarray}
& &\hspace*{-10mm}
F_{ijk} (\alpha_{j,1}x_{k,1} - \alpha_{j,0}^{+}x_{k}) = 0 \ts,
\label{eq:F_ijk-alpha_d1} \\
& &\hspace*{-10mm}
F_{ijk} (\alpha_{j,1}^{\ast}y_{k,1} - \alpha_{j,0}^{+}y_{k}) = 0 \ts,
\label{eq:F_ijk-alpha_d2} \\
& &\hspace*{-10mm}
F_{ijk} (\alpha_{j,1}y_{k} + \alpha_{j,0}^{-}y_{k,1}) = 0 \ts,
\label{eq:F_ijk-alpha_d4} \\
& &\hspace*{-10mm}
F_{ijk} (\alpha_{j,1}^{\ast}x_{k} + \alpha_{j,0}^{-}x_{k,1}) = 0 \ts,
\label{eq:F_ijk-alpha_d5} 
\end{eqnarray}
The symmetry of the equation under;
\begin{eqnarray*}
(x_{j},\ts x_{j,1}) \to (y^{\ast}_{j},\ts y^{\ast}_{j,1}) \ts,
\end{eqnarray*}
makes it possible to set $(y^{\ast}_{j},\ts y^{\ast}_{j,1})
= (\theta x_{j},\ts \theta x_{j,1})$. Then we have,
\begin{eqnarray*}
\gamma_{i}^{0+} \nts\nts&=&\nts\nts
2F_{jik} (1 + |\theta|^2) x_{j} x^{\ast}_{k} \ts, \\
\gamma_{i}^{0-} \nts\nts&=&\nts\nts
2F_{jik} (1 + |\theta|^2) x_{j,1} x^{\ast}_{k,1} \ts, \\
\gamma_{i,1}    \nts\nts&=&\nts\nts
2F_{jik} (1 + |\theta|^2) x^{\ast}_{j,1} x_{k} \ts, 
\end{eqnarray*}
It can be shown in this case that,
\begin{eqnarray*}
|\alpha_{0}| \nts\nts&=&\nts\nts
2\pi^2 \eta(\tau)^{4} \ts, \\
|\alpha_{1}| \nts\nts&=&\nts\nts
2\pi^2 \eta(\tau_1)^{4} \ts, 
\end{eqnarray*}
where $\eta(\tau)$ is the Dedekind modular form. Instead of
(\ref{eq:dfm_eom_clt_1}) and (\ref{eq:dfm_eom_clt_2}) we have,
\begin{eqnarray}
4\eta(\tau)^{12} \rho^{\pm}_{i} + 4\eta(\tau_{1})^{8}
\eta(\tau)^{4}\sigma^{\pm}_{i,1}
\nts\nts&=&\nts\nts \frac{\ds 1}{\ds 8\pi^{6}}
g \gamma^{\pm}_{i} \ts, 
\label{eq:ins_eom_clt_1} \\
4\eta(\tau_{1})^{12} \sigma_{i;1,1} + 4\eta(\tau)^{8}
\eta(\tau_{1})^{4}\rho_{i,1}
\nts\nts&=&\nts\nts \frac{\ds 1}{\ds 8\pi^{6}}
g \gamma_{i,1} \ts, 
\label{eq:ins_eom_clt_2}
\end{eqnarray}

One might think that it may be possible to get a one-brane equation by
putting
\begin{eqnarray*}
A^{2} = A^{3} = 0 \es\mbox{and}\es x_{j,1} = y_{j,1} = 0 \es
\mbox{in the above equations.}
\end{eqnarray*}
But, unfortunately, we can easily prove
that the resulting set of equations do not have a solution. One-brane
case must be defined in a different way. Another case is that of M2-brane.
We write down the equations when we have the extended Pauli spin.
\begin{eqnarray}
& &\hspace*{-10mm}
\vec{\alpha}^{0+} \times \vec{x} 
+ \vec{A}^{11} \times \vec{\bar{x}} = 0 \ts,
\label{eq:ePauliS_1} \\
& &\hspace*{-10mm}
\vec{\alpha}^{0+} \times \vec{\bar{x}} 
+ \vec{A}^{11} \times \vec{x} = 0 \ts,
\label{eq:ePauliS_2} \\
& &\hspace*{-10mm}
\vec{\alpha}^{0+} \times \vec{x}_{a,b}
+ \vec{A}^{11} \times \vec{\bar{x}}_{a,b} = 0 \ts,
\label{eq:ePauliS_3} \\
& &\hspace*{-10mm}
\vec{\alpha}^{0+} \times \vec{\bar{x}}_{a,b}
+ \vec{A}^{11} \times \vec{x}_{a,b} = 0 \ts,
\label{eq:ePauliS_4}
\end{eqnarray}
\begin{eqnarray}
& &\hspace*{-10mm}
2\vec{\alpha}^{0+} \times (\vec{\alpha}^{0+} \times \vec{\alpha}^{0-})
+ 4\vec{A}^{11} \times (\vec{\alpha}^{0+} \times \vec{A}^{11}) \nonumber \\
& & +\ts 2\kappa^2\bigl\{ (\vec{\alpha}^{0+} \times \vec{\alpha}^{0-}
\cdot \vec{I}) \vec{\alpha}^{0+} \times \vec{I}
+ (\vec{\alpha}^{0+} \times 
\vec{A}^{11} \cdot \vec{I}) \vec{A}^{11} \times \vec{I} \bigr\}
= 4ig(\vec{x}^{\ast}_{a,b} \times \vec{x}_{a,b}) \ts,
\label{eq:ePauliA_1} \\
& &\hspace*{-10mm}
2\vec{\alpha}^{0-} \times (\vec{\alpha}^{0-} \times \vec{\alpha}^{0+})
+ 4\vec{A}^{11} \times (\vec{\alpha}^{0+} \times \vec{A}^{11}) \nonumber \\
& & +\ts 2\kappa^2\bigl\{ (\vec{\alpha}^{0-} \times \vec{\alpha}^{0+}
\cdot \vec{I}) \vec{\alpha}^{0-} \times \vec{I}
+ (\vec{\alpha}^{0-} \times 
\vec{A}^{11} \cdot \vec{I}) \vec{A}^{11} \times \vec{I} \bigr\}
= - 4ig(\vec{\bar{x}}^{\ast}_{a,b} \times \vec{\bar{x}}_{a,b}) \ts,
\label{eq:ePauliA_2} \\
& &\hspace*{-10mm}
2\vec{\alpha}^{0-} \times (\vec{A}^{11} \times \vec{\alpha}^{0+})
- 2\vec{\alpha}^{0+} \times (\vec{\alpha}^{0-} \times
\vec{A}^{11}) \nonumber \\
& & +\ts 2\kappa^2\bigl\{ (\vec{\alpha}^{0-} \times \vec{\alpha}^{0+}
\cdot \vec{I}) \vec{\alpha}^{0-} \times \vec{I}
+ (\vec{\alpha}^{0-} \times 
\vec{A}^{11} \cdot \vec{I}) \vec{A}^{11} \times \vec{I} \bigr\}
= ig(\vec{\bar{x}}^{\ast}_{a,b} \times \vec{x}_{a,b}
+ \vec{x}^{\ast}_{a,b} \times \vec{\bar{x}}_{a,b}) \ts.
\label{eq:ePauliA_3}
\end{eqnarray}
Equations (\ref{eq:ePauliS_1})-(\ref{eq:ePauliS_4}) should be coupled with
the equations (\ref{eq:ePauliA_1})-(\ref{eq:ePauliA_3}).
%
%
\section{Gravity}\label{sec:Gravity}
We might define the gravitational field (metric tensor) as a functional of
the space-time coordinate. But by doing so we must introduce a new action
to determine the form of the functional. This is in a way inappropriate 
since we want the original matrix action to be able to determine everything.

  An elementary way to define the metric tensor is the following:
First we define the operator $\nabla$ which operates on any function defined
on the space of clusters by,
\begin{eqnarray}
\nabla f(n) = f(n) - f(n-1) \ts. 
\label{def:nablaOnF}
\end{eqnarray}
Then, 
\begin{eqnarray}
ds(n)^2 = g^{\rm v}_{\mu,\nu}(n) T_{r}
\bigl( \nabla A^{\mu}(n) \nabla A^{\nu}(n) \bigr) = T_{r}
\bigl( g_{\mu,\nu}(n) \nabla A^{{\rm v},\mu}(n)
\nabla A^{{\rm v},\nu}(n) \bigr) \ts. 
\label{def:metricOnF}
\end{eqnarray}
Here the notation ${\rm v}$ stands for the vacuum.
The vacuum solution $A^{{\rm v},\mu}(n)$ which corresponds to the solution
$A^{\mu}(n)$ can be defined by taking the limit $g \to 0$ and 
$f_{ijk} \to 0$ in the solution $A^{\mu}(n)$. The latter limit can be taken
by multiplying a small parameter to $f_{ijk}$ and taking it to zero.

The general covariance is defined as the transformation property of
$g$ under the interchange of clusters: For
$n^{\prime} = n^{\prime}(n)$, we have,
\begin{eqnarray}
T_{r} \bigl( g^{\prime}_{\mu,\nu}(n^{\prime}) \nabla
A^{{\rm v},\mu}(n^{\prime}) \nabla A^{{\rm v},\nu}(n^{\prime}) \bigr)
= T_{r} \bigl( g_{\mu,\nu}(n) \nabla A^{{\rm v},\mu}(n)
\nabla A^{{\rm v},\nu}(n) \bigr) \ts.
\label{def:genCovOnClstr}
\end{eqnarray}
The vacuum metric $g^{\rm v}_{\mu,\nu}$ could be just the flat metric
$\eta_{\mu,\nu}$, but it could also be a metric of the de\,Sitter or
the anti-de\,Sitter space. Either way, we assume it is proportional to
the unit matrix.

The connection is defined as
\begin{eqnarray}
A_{\mu\nu,\rho\kappa} = \frac{\ds 1}{\ds 4} \left[\ts\bigl\{ g_{\mu\nu}(n)
\nabla g_{\rho\kappa}(n) - g_{\rho\nu}(n) \nabla g_{\mu\kappa}(n) \bigr\}
- \bigl\{ \mu,\nu \leftrightarrow \rho,\kappa \bigr\} \ts\right] \ts.
\label{def:cnnctnClstr}
\end{eqnarray}
The justification comes from the idea of non-commutative
geometry\,\cite{ConneMar}. In this particular case, we treat $\nabla$ as
the Dirac operator defined on the functional space of $g_{\mu\nu}$.
We have the following identities,
\begin{eqnarray*}
A_{\mu\nu,\rho\kappa}(n) = A_{\rho\kappa,\mu\nu}(n) \ts, \quad
A_{\mu\nu,\rho\kappa}(n) = - A_{\rho\nu,\mu\kappa}(n) 
= A_{\mu\kappa,\rho\nu}(n) \ts.
\label{eq:cnnctnIds}
\end{eqnarray*}
The curvature is defined again loosely following non-commutative
geometry,
\begin{eqnarray}
R_{\mu\nu,\rho\kappa}(n) \equiv \nabla A_{\mu\rho,\nu\kappa}(n) + 
\frac{\ds 1}{\ds 2} \bigl( A_{\mu\rho,\nu^{\prime}\kappa^{\prime}}(n)
A^{\nu^{\prime}\kappa^{\prime}}_{\quad\nu\kappa}(n) -
A_{\nu\rho,\nu^{\prime}\kappa^{\prime}}(n)
A^{\nu^{\prime}\kappa^{\prime}}_{\quad\mu\kappa}(n) \bigr) \ts.
\label{def:curvatureNCG}
\end{eqnarray}
We have the following identities,
\begin{eqnarray*}
R_{\mu\nu,\rho\kappa}(n) = - R_{\nu\mu,\rho\kappa}(n)
= - R_{\mu\nu,\kappa\rho}(n)\ts.
\label{eq:curvatureIds}
\end{eqnarray*}
We do not have the cyclic identity for $R_{\mu\nu,\rho\kappa}(n)$.
We can, therefore, define the Ricci tensor uniquely but it is not 
necessarily symmetric under the interchange of suffices. The vacuum 
curvature is assumed to satisfy (neglecting the symbol ${\rm v}$),
\begin{eqnarray}
R_{\nu\kappa}(n) = g^{\mu\rho}(n) R_{\mu\nu,\rho\kappa}(n)
= \lambda g_{\nu\kappa}(n) \ts.
\label{def:ricciTnsr}
\end{eqnarray}
This equation can be easily solved by the following anzatz,
\begin{eqnarray*}
\nabla g_{\mu\nu}(n) = g_{\mu\nu}(n) - g_{\mu\nu}(n-1) 
= \delta g_{\mu\nu}(n) \ts,
\label{def:metricAnzatz}
\end{eqnarray*}
where $\delta$ is a certain rational number. The $g_{\mu\nu}(n)$ can
easily be solved,
\begin{eqnarray}
g_{\mu\nu}(n) = \frac{\ds 1}{\ds (1-\delta)^{n-1}} g_{\mu\nu}(1) \ts.
\label{def:solMetAnzatz}
\end{eqnarray}
Equation (\ref{def:ricciTnsr}) is reduced to,
\begin{eqnarray}
\lambda = \frac{\ds N-1}{\ds 8} \left[ (N+6)\delta^{2}
-4\delta^{3} \right] \ts.
\label{eq:redVacCurvature}
\end{eqnarray}
$N$ is the dimension of the space-time.

These equations give a new interpretation to the cosmological constant
$\lambda$ as follows: When $\lambda$ is small we have,
\begin{eqnarray}
g_{\mu\nu}(n) = \exp\left\{ (n-1)\delta \right\} g_{\mu\nu}(1) \ts.
\label{eq:smallDeltaInt}
\end{eqnarray}
Assuming that $\delta$ is a negative number, this equation shows that
$1/|\delta|$ is the effective number of vacuum space-time points.
The distance between the points with the cluster number larger than
$1/|\delta|$ becomes small compared to the Planck scale and so the points
become undistinguishable.

In the case of our universe, $\lambda \approx 10^{-120}$ which gives
$1/|\delta| \approx 10^{60}$. The curvature size of the universe is
$10^{28}\,{\it cm}$ and if we divide this by the Planck length 
$10^{-33}\,{\rm cm}$, we get $10^{61}$. Since our space-time is 
4-dimensional we have $10^{244}$ as the number of points of the current
universe. The corresponding vacuum seems to have only one-dimensional
degrees of freedom.
\vspace*{5mm}
%
\section{(First) Quantization and the Non-commutative Geometry}
\label{sec:1stQiztnNCG}
The matrix model is equipped with its algebra and the Hilbert space as the
representation space of the algebra. The (first) quantization can be done
utilizing this Hilbert space. The missing operator is the canonical
conjugate variable $P_{\mu}$ to $A^{\mu}$ and $Q_{\alpha}$ to
$\Psi_{\alpha}$ with the commutation relation,
\begin{eqnarray*}
[ P_{\mu},\ts A^{\nu} ] = -i \delta^{\nu}_{\mu} \ts,
\label{eq:commRltn-PA}
\end{eqnarray*}
and
\begin{eqnarray*}
\{ \Psi_{\alpha},\ts \overline{Q}_{\beta} \}
= \{ \overline{\Psi}_{\alpha},\ts Q_{\beta} \} =
i \delta_{\alpha\beta} \ts.
\label{eq:anticommRltn-PsiQ}
\end{eqnarray*}
If we interpret the algebra and the Hibert space to be that of
non-commutative geometry, the missing operator to complete
the Conne's triple spectra\,\cite{JBNL} is the Dirac operator. 
It is natural to assume that
they are also given by $P_{\mu}$ and $Q_{\beta}$.
In this interpretation, the solution of the equations we have been discussing
is not a classical solution but a quantum one with $A^{\mu}$ and
$\Psi_{\alpha}$ diagonalized up to the clusters.

We can define the gauge fields in the space of matrix algebra using the standard prescription of non-commutative geometry\,\cite{JBNL}. The relation of
gravity defined in the last section and the one defined in this way is
not clear at this time.

$P_{\mu}$ and $Q_{\beta}$ have the other roles. The former is the operator
for the translational invariance of the Lagrangian (\ref{eq:actionIKKT}),
\begin{eqnarray*}
A^{\mu} &\longrightarrow& A^{\mu} + a^{\mu} I \ts, \\
\Psi    &\longrightarrow& \Psi \ts.
\label{eq:transInv-APsi}
\end{eqnarray*}
The latter is the operator for one of the supersymmetry transformaions
defined in\,\cite{IKKT},
\begin{eqnarray*}
\Psi    &\longrightarrow& \Psi + \xi I \ts, \\
A^{\mu} &\longrightarrow& A^{\mu} \ts.
\label{def:eq:transInv-IKKT}
\end{eqnarray*}
This corresponds to the second supersymmetry defined by IKKT as
$\delta^{(2)}$\,\cite{IKKT}.

The supersymmetry operator corresponding to $\delta^{(1)}$ of IKKT can
be defined as follows,
\begin{eqnarray*}
Q^{(1)}_{\alpha}
\nts\nts&=&\nts\nts (\Gamma_{\mu}\Psi)_{\alpha} P^{\mu}
- \frac{\ds 1}{\ds 2} [A_{\mu},\ts A_{\nu}]
(\Gamma^{\mu\nu} Q)_{\alpha} \ts, \\
\overline{Q}^{(1)}_{\alpha}
\nts\nts&=&\nts\nts - (\overline{\Psi}\Gamma_{\mu})_{\alpha} P^{\mu}
- \frac{\ds 1}{\ds 2} [A_{\mu},\ts A_{\nu}]
(\overline{Q}\Gamma^{\mu\nu})_{\alpha} \ts.
\label{def:susyOP-IKKT }
\end{eqnarray*}
We the get the following anti-commutation relations for these operators,
\begin{eqnarray*}
& &\hspace*{-10mm}
\{ Q^{(1)}_{\alpha},\ts Q^{(1)}_{\beta} \}
= \{ \overline{Q}^{(1)}_{\alpha},\ts \overline{Q}^{(1)}_{\beta} \}
= 0 \ts, \\
& &\hspace*{-10mm}
\{ \overline{Q}^{(1)}_{\alpha},\ts Q^{(1)}_{\beta} \}
= P^{\rho} \ts [A_{\mu},\ts A_{\nu}]
(\delta^{\nu}_{\rho}P^{\mu} - \delta^{\mu}_{\rho}P^{\nu}) \ts.
\label{def:antiCommRel-IKKT }
\end{eqnarray*}
Defining the gauge fields including the gravity using Dirac operators
$P^{\mu}$ and $Q_{\alpha}$ will be discussed in a future publication.
\vspace*{5mm}
%
\section{Conclusions}\label{sec:Conclusions}
The proposal is made that the space-time is constructed out of integers
or rational numbers rather than being a continuum. This is based on taking
the matrix model as the fundamental theory of space-time as it is. The
proposal is substantiated by observing that the reduced matrix model
equations can be thought of as the elliptic curves which play the central
role in number theory. The whole machinery of the number theory developed
over the centuries could be incorporated into the understanding of our
space-time.

  The space-time coordinates $A_{\mu}$ and $\Psi_{\alpha}$ are
mutually non-commuting but we assume that they can be partially
diagonalized up to certain clusters. We are then left with the equations
of finite sized matrices and they form a system of coupled elliptic curves.

  The metric tensor for this space-time can be calculated using the
solution and the prescription described in section\,\ref{sec:Gravity}
The gravity is
completely determined by the original reduced matrix model equations in
this sense. It is necessary to define the vacuum metric for this purpose.
It could be just a flat metric but it could also have a finite cosmological
constant when the number of effective space-time points is finite.
The latter is given by the inverse of the former thus providing a new
meaning to the cosmological constant. The fact that our space-time seems
to have a finite cosmological constant corresponds to the fact that it is
composed of finite number of effective space-time points.

  The (first) quantized system of the matrix model can be understood
within the framework of non-commutative geometry. The fact that the
matrix model is equipped with the algebra and the representation space
allows us to introduce a canonical variable as the Dirac operator in the
non-commutative geometry. This possibly provides us an alternative way
to define the gravity but it starts with the level of a connection or a
gauge field. Our naive definition of the metric tensor must be related
to the latter definition in a certain way.

  Non-commutative geometry itself is very much related to the number
theory but it is outside of the scope of the present work.

  Finally I will list some of the problems for the future
investigations:
\begin{itemize}
\item[(1)] Can the non-commutative definition of gauge theory lead
to the $E(8) \times E(8)$ theory?
\item[(2)] The discussions of section\,\ref{sec:Preparations}
suggest that we may need the non-abelian Galois extension of the field.
What is the physical meaning of the Galois group?
\item[(3)] Is it possible to understand our assertion that equation
(\ref{eq:triProduct}) should be modular within the framework of
Langlands philosophy?
\item[(4)] Is it possible to incorporate all the geometrical concept
related to the string theory into the number theory formulation of
the matrix model?
\end{itemize}
\vspace*{1.8cm}

\noindent{\large \bf Acknowledgment} \\
I would like to thank H.~Hagura for the technical assistance.
\vspace*{1.8cm}
\end{document}